\documentclass[12pt,preprint]{aastex}

\begin{document}

\title {Six White Dwarfs with Circumstellar Silicates}

\author{M. Jura\footnote{Department of Physics and Astronomy and Center for Astrobiology, University of California, Los Angeles CA 90095-1562; jura, ben @astro.ucla.edu}, J. Farihi,\footnote{Department of Physics and Astronomy, University of Leicester, Leicester LE1 7RH, United Kingdom; jf123@star.le.ac.uk}\,\,\,\,\&\, B. Zuckerman$^{1}$}

\begin{abstract}
{\it Spitzer Space Telescope} spectra  reveal 10 ${\mu}$m silicate emission from circumstellar dust orbiting six externally-polluted white dwarfs.   Micron-size glasses with an olivine stoichiometry  can account for  the distinctively broad wings that extend  to 12 ${\mu}$m; these particles likely are produced by tidal-disruption of asteroids. The  absence of infrared PAH features is consistent with a scenario where extrasolar rocky planets are assembled from
 carbon-poor solids.
\end{abstract}

\section{INTRODUCTION}

Approximately  1-3\%  of single white dwarfs with cooling ages less than 0.5 Gyr possess an infrared excess resulting from a circumstellar disk (Farihi et al. 2008b). These same stars also have distinctively high atmospheric calcium abundances (Kilic et al. 2006, Jura et al. 2007a) even though  photospheric metals normally are absent in white dwarfs in this temperature range (Zuckerman et al. 2003).  The disks likely are caused by the tidal disruption of asteroids, and there is subsequent metal accretion  onto the white dwarfs  (see Jura 2008).  Polluted white dwarfs and their disks are  powerful tools to investigate extrasolar asteroids -- the building blocks of extrasolar rocky planets.

There are
14 single white dwarfs currently  known to have a definite or highly-likely continuum infrared excess (Zuckerman \& Becklin 1987, Becklin et al. 2005, Kilic et al. 2005, 2006, Jura et al. 2007a, Kilic \& Redfield 2007,  von Hippel et al. 2007, Farihi et al. 2008a,b, Brinkworth et al. 2008, Melis et al. 2008, in preparation).  Previously, 
spectra with the Infrared Spectrograph (IRS, Houck et al. 2004) on the {\it Spitzer Space Telescope} (Werner et al. 2004) have been reported for two stars: G29-38 (Reach et al. 2005, Reach et al. 2008) and GD 362 (Jura et al. 2007b). 
Both stars exhibit a strong 10 ${\mu}$m emission feature whose red wing can be modeled as arising
from olivine-like glasses.   Here, we report IRS results for 6 additional white dwarfs with an infrared excess.

\section{OBSERVATIONS AND DATA}

IRS spectra would be desirable for all  white dwarfs with a definite or highly-likely continuum infrared excess.   We observed  the first 6 white dwarfs listed in Table 1.  An IRS spectrum  was obtained for LTT 8452 by another group but never published; for completeness, we report the results here.
We did not target G166-58 because  a nearby bright background galaxy precludes useful observations at ${\lambda}$ $>$ 8 ${\mu}$m.  SDSS 1228+1040 (Gaensicke et al. 2006, Brinkworth et al. 2008), PG 1457$-$086 (Farihi et al. 2008b) and Ton 345 (Melis et al. 2008, in preparation),  were not known to have an infrared excess at the time the present program was implemented.  

Observations with  IRS  were 
executed during {\em Spitzer} 
Cycle 4, between 2007 July and 2008 February.  Spectroscopy was performed in 
staring mode using modules SL1 and SL2 which essentially cover the 5--15 ${\mu}$m
region with spectral resolution between 64 and 128.    The respective ramp times for 
these targets in each module are listed in Table 1 and were constrained by the expected sky backgrounds in the region of each 
target.  Longer ramp times for these faint sources were prohibited 
by the predicted backgrounds lest the data become problematic due to saturation 
in the peak-up sections of the array ({\it Spitzer} Science Center 2007).  Additionally, IRS observations 
of the white dwarf LTT 8452, performed during Cycle 2 in 2006 October, 
were extracted from the {\em Spitzer} archive for analysis.

The spectral data were processed with the IRS calibration pipeline, versions 15 
(LTT 8452), 16 (EC 11507$-$1519 and GD 56), and 17 (GD 16, GD 40, GD 133, and PG 1015+161).
The combined, sky-subtracted frames were manually corrected for bad pixels using
the IRSCLEAN\footnote{http://ssc.spitzer.caltech.edu/postbcd/irsclean.html} package, 
then processed with SPICE\footnote{http://ssc.spitzer.caltech.edu/postbcd/spice.html}
to perform spectral extraction.  The {\sf optimal} extract function was used for 
these relatively low signal-to-noise (S/N) data; first with the default aperture 
function (8 pixels at 12 $\mu$m) to assess the proper calibration level, then with a 
custom aperture function (4 pixels at 12 $\mu$m) to produce a higher S/N extraction.  
Data at both nod positions were averaged and the custom extraction data were 
scaled to the well-calibrated level of the default extraction, with all orders 
combined and averaged in regions of overlap.  No inter-order adjustments were 
made to the data.

The S/N can be estimated by examining 
the spectra themselves and evaluating the deviation within regions where a
featureless, flat continuum might be expected.  In this manner, the S/N over
the 9--11 $\mu$m region was estimated by taking the mean flux in that region
and dividing by the standard deviation in the 6-8 $\mu$m region.  The spectra
shown in Figures 1--6 have S/N between 3.5 and 8.0 in the
9--11 $\mu$m interval.  

The IRS dataset for LTT 8452 is somewhat problematic; it appears to suffer
from saturation effects in the peak-up portions of the array due to high background,
and possibly from point source overexposure.  While the latter does not produce an
unrecoverable problem for IRS data, the former has the potential to significantly 
compromise the S/N and the spectrophotometric flux calibration.  In the co-added
spectroscopic BCD frames, the median flux levels in the blue and red sub-arrays 
are just above 850 and 1050 electrons s$^{-1}$, respectively, corresponding to point
source flux levels of more than 0.1 Jy, a value which saturates the blue sub-array in a
ramp time of 14 seconds, and the red sub-array in around 30 seconds ({\it Spitzer} Science Center 2007).
At least two bright point sources are seen in the peak-up arrays 
in the nod-subtracted spectroscopic frames, and it is likely the overall background 
level exceeded the recommended 25 MJy/sr limit for 240 sec ramp times (consistent
with the background predictions made by SPOT for this object).  In any case the 
spectroscopic signal is unaffected by flux bleeding from the peak-up sub-arrays; hence one can confidently say there are no features in the
spectrum of LTT 8452 which are the result of the overexposure.  However, the S/N
is clearly compromised relative to expectations, and the overall calibration level
does not match its IRAC photometry.  The spectrum of LTT 8452 has been re-scaled up 
by a factor of 2.5 to match these fluxes.

No detectable signal was present in the IRS observations of PG 1015+161.
It is likely the source was too faint for the short ramp times required by the
relatively high background region of the sky.

The resulting spectra with both smoothed and unsmoothed data along with relevant photometry are displayed in Figures 1--6.   Additionally, for each star, we present  the profile of the interstellar silicate feature inverted from absorption to emission  (Kemper et al. 2004),  the profile of the silicate emission in  BD +20 307, a main-sequence debris disk star with a remarkably strong 10 ${\mu}$m feature (Song et al. 2005), and a model of the circumstellar disk emission described in more detail below.  According to Min et al. (2007), the interstellar silicate profile of Kemper et al. (2004) is generally representative of diffuse interstellar matter.  In the study by Wyatt et al. (2007) of main-sequence debris disks with warm dust that emits near 10 ${\mu}$m, BD +20 307 has by far the largest fraction of stellar luminosity that is reprocessed by dust in a system
where the grains cannot be primordial.  Thus the infrared spectrum of BD +20 307 likely  is our best current measure of the spectrum of dust
derived from extrasolar rocky parent bodies.  Rhee et al. (2008) have reported an infrared excess around HD 23514, an F-type main-sequence star in the Pleiades, which, like BD +20 307, also reprocesses more than 1\% of the host star's luminosity.  However, the silicate emission from HD 23514 peaks near 9 ${\mu}$m,  and the feature is very
non-standard. Each observed white dwarf displays silicate emission that peaks near 10 ${\mu}$m. 
All  profiles are notably better matched by the profile from BD +20 307  than the interstellar profile, and in all cases the red wing of the feature extends to at least 12 ${\mu}$m.  The white dwarf spectra are more closely matched by dust derived from rocky parent bodies than by interstellar dust. 

For comparison, the smoothed data  from Figures 1--6 plus the IRS data for GD 362 and G29-38 are shown in Figure 7.   The features are not identical; GD 362 exhibits the greatest line-to-continuum ratio while GD 133 possesses a particularly broad feature.  We see no evidence of PAH emission or any other spectroscopic features.

\section{MODELS}

The continuum infrared fluxes which typically reprocess ${\sim}$1\% of a white dwarf's luminosity  are fit by
 a flat, passive, opaque dusty disk (Jura 2003, Jura et al. 2007a).   Our IRS observations show that typically ${\sim}$0.1\% of the stellar
 luminosity is reprocessed as a 10 ${\mu}$m silicate emission.  While  the overall disk parameters are largely determined by the continuum fluxes,  
we use the IRS data to constrain the composition of the circumstellar dust.

To exploit the IRS data, we elaborate upon our previous very simple models for white dwarf dust disks.  In these models,   the inner boundary is near the location where sublimation of silicates becomes rapid while the outer boundary lies within the tidal radius of the star.   These simple  models presume that the disk is vertically isothermal at every radius, as may occur in regions where the gas density is high enough for thermal conduction to be important and emission from such regions cannot explain the silicate features.  There are at least two possible generalizations.
  (1) The ``top" of the disk is directly illuminated by the light from the host star and is hotter than
  the midplane if thermal conduction is not dominant.  Disks with little gas possess a  vertical temperature gradient 
  and an emission feature can result. (2)  The outer zones of the disks can be optically thin.  To explain
  the data for GD 362 with its particularly strong silicate emission, Jura et al. (2007b) employed both
  elaborations to the simple disk in a model with three radial zones. Here, except for GD 56 described below, we match the 
  data in Figures 1--6  with models consisting of two radial zones.

We have found that both kinds of   generalizations to the simple disk model can be made to fit the data. In order to compute models
with a hot layer on top of a cold interior, we need to estimate  thermal conduction by the gas as a function of location in the disk.  While at least some disks contain gas (Gaensicke et al. 2006, 2007), its amount and  spatial distribution are unknown. 
In this paper, because fewer unknown parameters are required,  we only present results for calculations with an optically thin outer region.

In our  models, 
each radial zone  is characterized by the Planck-mean infrared optical depth measured vertically through the disk, ${\tau}$.  In the inner  zone we assume ${\tau}$ $>>$ 1 and that  the dust is vertically isothermal.   In this zone, at distance
$R$ from the star of radius $R_{*}$ and effective temperature $T_{eff}$, the disk temperature, $T$, is (Jura 2003):
\begin{equation}
T\;=\;\left(\frac{2}{3{\pi}}\right)^{1/4}\left(\frac{R_{*}}{R}\right)^{3/4}\,T_{eff}
\end{equation}
The observed flux from the opaque portion of the disk depends upon its inclination angle.

In the outer zone,  we take ${\tau}$ $<<$ 1.  However, even though the disk
is vertically transparent, since light from the incident star arrives nearly parallel to the disk surface,  it is mostly absorbed.
If ${\alpha}$ denotes the ratio of the mean opacity for absorption of optical and ultraviolet light compared to the mean opacity at infrared wavelengths where the dust emits,
then at the ``top" of the disk where the vertical optical depth is zero, the temperature, $T(0)$, is (Jura et  al. 2007b):
\begin{equation}
T(0)\;=\;\left(\frac{{\alpha}}{8}\right)^{1/4}\,\left(\frac{R_{*}}{R}\right)^{1/2}\,T_{eff}
\end{equation}
For both the relatively large grains and  the temperature range of interest here, we adopt the very rough approximation that ${\alpha}$ = 1.  Therefore, in this  zone, the vertical temperature variation is (Jura et al. 2007b):
\begin{equation}
T({\tau})\;{\approx}\;T(0)\,\left(e^{-{\tau}/{\mu}_{0}}\right)^{1/4}
\end{equation}
In equation (3), ${\mu}_{0}$ is the mean cosine of the slant angle of the radiation incident from the star onto the disk (Jura et al. 2007b):
\begin{equation}
{\mu}_{0}\;=\;\frac{4}{3\,{\pi}}\,\frac{R_{*}}{R}
\end{equation}
Because we take ${\tau}$ $<<$ 1, emission from both ``sides" of the disk
contributes to the observed flux which is independent of the disk inclination angle.

We list in Table 2, our adopted model parameters for each star. As described below, the model for GD 56 is more complex and requires additional parameters.  Each model requires two stellar parameters: temperature, $T_{eff}$ and angular radius, $R_{*}/D$ where $D$ is the distance between the Sun and the white dwarf.  There are also 5 disk parameters:
inner disk radius, $R_{in}$, transition radius from being optically thick to optically thin, $R_{trans}$, outer disk radius, $R_{out}$, disk inclination angle, $i$, and grain radius, $a$.  
We employ values of $T_{eff}$ derived from the previous studies listed in Table 2.  We  fit $R_{*}/D$ from
the $J$ and $H$ photometry with the assumption that emission from the disk is unimportant at these
wavelengths and by approximating  the stellar spectrum as a blackbody at its effective temperature.  The disk parameters are chosen to  fit  the data shown in Figures 1--6.

  We find the  inner boundaries of the dust disks to be located where the grain temperature ranges between 1000 K and 1500 K according to equation (1) and therefore solids rapidly sublimate.    We take $R_{outer}$ to fit the strength of the silicate emission and to lie within the tidal radius of the white dwarf. 
Following  Jura et al. (2007b), we assume the particles are glassy silicate spheres with an olivine stoichiometry  with equal abundances of magnesium and iron (Dorschner et al. 1995).  Olivines are  the dominant silicates in a model of the  spectrum of G29-38 (Reach et al. 2005), while Lisse et al. (2008)   found that  the red wing of the 10 ${\mu}$m feature in the spectrum of the debris-disk binary system HD 113766 is mainly the result of olivines.
For simplicity, we assume single-size dust particles with  $a$ between 1 ${\mu}$m and 2 ${\mu}$m. In order to reproduce a strong emission feature at wavelength, ${\lambda}$, we expect that $a$ $<$ ${\lambda}/(2{\pi}$.  This provides an upper limit to the particle size.  We infer a lower bound to the particle size from two arguments.  Small grains are likely to be relatively hot and sublimate rapidly.  Also, as shown by Dorschner et al. (1995), particles with radii larger than 1 ${\mu}$m display enhanced red wings to the silicate feature as found in our data.  The model disk need not be unique (Jura et al. 2007a), but, here, we present results for only one set of parameters.
Results  are shown in Figures 1--6; within the uncertainties, the  models reproduce the data.   

Jura et al. (2007a) found their flat disk models  for GD 56 under-predicted the measured fluxes between 3.6 ${\mu}$m and 7.9 ${\mu}$m.    Since the infrared spectrum of GD 56 displays silicate
emission similar to that found in other white dwarf disks, dust reprocessing of light from the central star -- rather than some other source in the system -- seems likely.  Here, we show that a passive, warped disk can account for the data.

Armitage \& Pringle (1997) have discussed warped disks, and Jura et al. (2007b) applied their models to a portion of the disk around GD 362.  Asymmetries in the double-peaked optical emission lines seen in two disks (Gaensicke et al. 2006, 2007) indicate deviations from flat systems with material in circular orbits.  Here, we consider a warped disk with  a three-radial-zone model.   The innermost zone of the  disk is flat, opaque, vertically-isothermal, and unwarped.  There is  a transition radius, $R_{warp}$, where the disk is also opaque and vertically-isothermal, but  warped. If ${\mu}_{warp}$ is the cosine of the angle between the warped portion the disk and the line of sight to the central star so that ${\mu}_{warp}$ = 0 corresponds to no warp, then in this portion of the warped disk:
\begin{equation}
T(0)\;=\;\left(\frac{{\mu}_{warp}}{2}\right)^{1/4}\,\left(\frac{R}{R_{*}}\right)^{1/2}\,T_{eff}
\end{equation}
We propose that the outer zone of the disk is optically thin but also warped.  With the assumption that ${\alpha}$ ${\approx}$ 1, the  temperature variation within the outer zone of disk is modified from equation (3) to:
\begin{equation}
T\;=\;T(0)\,\left(e^{-{\tau}/{\mu}_{warp}}\right)^{1/4}
\end{equation}
In this warped disk case, only the top half of the system is illuminated even in the optically thin outer zone.  We consider a parameter regime where the warp angle is sufficiently small that the solid angle subtended by
the infrared-emitting disk is not appreciably changed.

The warped portions of the disk receive more light and are warmer
than if the disk was flat. Consequently, the model can account for  more near-infrared emission than
predicted for a flat disk.
We show in Figure 3, with the warping parameters listed in the caption,  a model that reproduces the data for GD 56. 

Above, we have described disk models for the infrared excess seen at six  white dwarfs.  A natural explanation for
such a disk is the tidal disruption of an asteroid that strayed close to the star.
An alternative model to explain circumstellar dust orbiting a white dwarf is that grains are accreted from the interstellar medium and then drift inwards under the action of Poynting-Robertson
drag.  Such a model can be  tested  for 
white dwarfs where hydrogen is the dominant light element since the dwell time of metals in the atmosphere  is sufficiently short that accretion must be ongoing.  Specifically, four white dwarfs in Table 4 have atmospheres where hydrogen is the dominant light element, and the
atmospheric dwell times for calcium are 10 days, 7 days, 9 days and 200 years, for GD 56, GD 133, EC 11507$-$1519 and LTT 8452, respectively (Koester \& Wilken 2006).   Using the metal accretion rates derived by Koester \& Wilken (2006), 
Jura et al. (2007a) showed that the measured value of F$_{\nu}$(24 ${\mu}$m) was lower than predicted by this interstellar accretion model  by  factors of 40 and 140 for GD 56 and GD 133, respectively.  Extending their analysis (equation [3] in Jura et al. 2007a) to LTT 8452,    and using the stellar distance and accretion rate given by  Koester \& Wilken (2006), the interstellar accretion model over-predicts 
 F$_{\nu}$(24 ${\mu}$m) by more than a factor of 1000.  There is no reported measurement of F$_{\nu}$(24 ${\mu}$) for EC 11507$-$1519, so we do not consider this star.  When relevant data are available, the tidally-disrupted asteroid model is favored.

 \section{DISCUSSION}
 
All eight white dwarfs with IRS-measured spectra of their infrared excess exhibit a 10 ${\mu}$m silicate emission feature with a red wing that extends beyond 12 ${\mu}$m (Reach et al. 2005, Jura et al. 2007b, Figures 1--6).  As discussed by Lisse et al. (2008),  such  features are characteristic of olivines and appear in the spectra of evolved solids associated with planet formation.
 
 All  white dwarfs with an infrared excess are sufficiently hot they could excite emission from  
circumstellar PAH's, yet none do (this paper, Farihi et al. 2008a, Jura et al. 2007b, Reach et al. 2005, Reach et al. 2008).  The mid-infrared spectra of 25\% of Herbig Ae stars with $T_{eff}$ near 10,000 K  are dominated by   PAH emission and they do not show any silicate features (Sloan et al. 2005).  At least 50\%   of Herbig  Ae  stars (Meeus et al. 2001) display at least some PAH emission even if not dominant. Thus,
qualitatively,  the absence of PAH emission in their IRS spectra suggests that  white dwarfs are orbited by silicate-rich, carbon-poor material.  Currently, GD 40 and GD 362,  two white dwarfs with infrared silicate emission, also have  a detailed measure of their atmospheric composition.  The apparently low carbon fraction in the disk is paralleled  with a low carbon abundance in their atmospheres  since $n$(C)/$n$(Fe) ${\leq}$ 1.02 in GD 362 (Zuckerman et al. 2007, Table 2) and
$n$(C)/$n$(Fe) = 0.1 in GD 40 (Wolff et al. 2002).  For reference, in the Sun, $n$(C)/$n$(Fe) = 9.3 (Lodders 2003).

The rocky material in the inner solar system is carbon deficient. In CI chondrites -- the most primitive of 
meteorities -- and in
Earth's mantle, $n$(C)/$n$(Fe) is 0.9 and 0.009, respectively  (Lodders 2003, McDonough \& Sun 1995).   Our new data are consistent with the view that extrasolar asteroids are carbon-deficient by a factor of 10 or more (Jura 2006, 2008).   Therefore the building blocks of extrasolar rocky planets have the same
substantial carbon deficiency as found in the inner solar system.
\section{CONCLUSIONS}

We report 10 ${\mu}$m silicate emission from dusty disks orbiting six  externally-polluted white dwarfs. The red wings extending to 12 ${\mu}$m are characteristic of micron-sized glasses with an olivine stoichiometry, particles that likely arise from tidally-disrupted asteroids.  The strong silicate emission and the absence of PAH emission support the scenario that extrasolar rocky planets are assembled from silicate-rich, carbon-poor solids.  

This work has been partly supported by NASA grants to UCLA and is based on observations made with the {\it Spitzer Space Telescope} which is operated by the Jet Propulsion Laboratory, California Institute of Technology, for NASA.

\begin{center}
{\bf APPENDIX}
\end{center}
\renewcommand{\theequation}{A\arabic{equation}}
\setcounter{equation}{0}
Reach et al. (2005)  modeled the infrared excess emission from  G29-38 with an optically thin circumstellar cloud including ${\sim}$3 ${\times}$ 10$^{17}$ g of silicates.  Here, we show that this mass of optically thin silicates is a natural extension of a 
flat, largely opaque disk described 
  in ${\S}$3.    The maximum vertical thickness, $Z_{abs}$, of any optically thin layer is determined by the cosine of the slant angle, ${\mu}_{0}$,  and the Planck-mean  opacity for absorption of this light, ${\chi}$ (cm$^{2}$ g$^{-1}$).  If the space density of the grains is ${\rho}$, then:
\begin{equation}
Z_{abs}\;=\;\frac{{\mu}_{0}}{{\rho}\,{\chi}}
\end{equation}
 With inner and outer radii such that $R_{out}$ $>>$ $R_{in}$, then using equation (4), the total mass of
optically thin dust, $M_{thin}$, on one side of the disk, is:
\begin{equation}
M_{thin}\;=\;{\int}_{R_{in}}^{R_{out}}\,{\rho}\,Z_{abs}\,2{\pi}\,R\,dR\;{\approx}\;\frac{8}{3}\,\frac{R_{*}\,R_{out}}{{\chi}}
\end{equation}
For spherical grains of matter density ${\rho}_{s}$ and radius $a$, we take:
\begin{equation}
{\chi}\;{\approx}\;\frac{3}{4{\rho}_{s}\,a}
\end{equation}
Reach et al. (2005) consider models with grains with a power law in the size distribution such that $dn/da$ varies as $a^{-3.5}$ and the radius range is  0.01 -- 1000 ${\mu}$m.  For such grains illuminated by white dwarfs with $T_{eff}$ near 10,000 K, a very approximate 
value of the mean size for absorption is $a$ near 1.0 ${\mu}$m.    In this case, with 
${\rho}_{s}$ ${\approx}$ 3  g cm$^{-3}$, then ${\chi}$ ${\approx}$ 2500 cm$^{2}$ g$^{-1}$ for optical and ultraviolet wavelengths  where the white dwarf emits most of its light. For $R_{out}$ ${\approx}$ 100 $R_{*}$ and with $R_{*}$ = 8.2 ${\times}$ 10$^{8}$ cm (Jura 2003), then  $M_{thin}$ ${\approx}$ 2 ${\times}$ 10$^{17}$ g.  Reach et al. (2005) estimate that the optically thin
cloud of dust contains ${\sim}$10$^{18}$ g of which 75\% is amorphous carbon to account for the continuum and 25\% is silicate to explain the 10 ${\mu}$m feature.  Their inferred mass of ${\sim}$2.5 ${\times}$ 10$^{17}$ g of optically-thin silicates is naturally computed for  a  flat, largely opaque, passive disk which then also accounts for the continuum fluxes.  

Reach et al. (2008) have acquired additional IRS data for G29-38 beyond that described in Reach et al. (2005), and they describe two models that account equally well for the data.  In their first model, the dust cloud is slightly opaque, while  their second model includes both a optically thick disk and an optically thin outer region composed of silicates.  This second model is similar to our models described above in $\S3$ for the six stars listed in Table 2. In this Appendix, we note that the optically thin
emission may reside ``on top" of the opaque disk rather than beyond it.  In any case, a major difference between the completely optically thin model and the model with both an optically thin region and a flat, opaque  disk and is the total amount of circumstellar mass. An optically
thin cloud has ${\leq}$ 10$^{18}$ g while an opaque disk can have orders of magnitude more mass.  The atmospheres of GD 40  and GD 362 which are helium-rich and therefore have a particularly deep convective zone, have been contaminated by ${\sim}$10$^{23}$ g of metals (Jura 2006, Zuckerman et al. 2007).  Therefore,  it is plausible that each of these two stars possesses a massive opaque disk in addition to  an optically thin dust cloud of ${\sim}$10$^{18}$ g.  

\newpage
\begin{center}
Table 1 -- Dusty White Dwarf IRS Targets
\\
\begin{tabular}{llllllllll}
\hline
\hline
WD & Name	&	F$_{\nu}$(2.2 ${\mu}$m) &	F$_{\nu}$(7.9 ${\mu}$m)	&Ramp &	No. Cycles &	Estimated	\\
&& & & Time &  & S/N  \\
&	& 			(mJy)	&	(mJy)	&	(sec)		&		\\
\hline
0146+187 &	GD 16	&	0.48	&	0.45	&	14	&	60	&	3.4 \\
0300-013 & GD 40	&	0.38	&	0.16	&	60	&	30	&	4.7 \\
0408$-$041 &GD 56 	&	0.58		&1.11 	&	60	&	7	&	8.0 \\
1015+161 &	PG 1015+161&	0.26	&	0.12	&	14	&	110	&	0 \\
1116+026&	GD 133	&	0.96	&	0.46 	&	14	&	60	&	6.0 \\
 1150$-$503 &	EC 11507$-$153&0.33	 &	\nodata	& 	60	&	7	&	3.7\\
2115$-$560 &	LTT 8452 &	1.64	&	0.88	&	240	& 	13	&	5.5 \\
\hline
\end{tabular}
\end{center}

\newpage
\begin{center}
Table 2 -- White Dwarf and Dust Disk  Properties 
\\
\begin{tabular}{lrllrrrcrr}
\hline
\hline
Star & $T_{eff}$ & $R_{*}/D$ & $\cos\, i$&  $R_{in}$ & $R_{trans}$ &$R_{out}$ & $a$ & References \\
  & (K) &  (10$^{-12}$)&  & ($R_{*}$)& ($R_{*}$) & ($R_{*}$)& (${\mu}$m)\\
\hline
GD 16 & 11,500 &     4.7 & 0.90 & 11 &23 &150 & 1.0 &(a)\\
GD 40 & 15,200 &   3.6 & 0.15 & 13 & 35& 100 & 1.5& (b)\\
GD 56 & 14,400 &  3.4 & 0.75 & 30 & 65 & 110 & 1.0 & (c)\\ 
GD 133 & 12,200 &7.0 & 0.19 & 12& 50&150 & 2.0 & (c)\\
EC 11507$-$1519 &    12,800 & 3.3& 1.00 & 10& 30& 150 & 1.5 & (c)   \\    
LTT 8452 &   9,700 & 11.0 & 0.27& 8 & 30&100 & 1.0 &(c)\\    
 
\hline
\end{tabular}
\end{center}
In this Table, $T_{eff}$, $R_{*}$ and $D$ are the stellar effective temperature, radius and distance, respectively and are taken from available measurements.  The disk inclination angle, $i$, (defined so that $i$ = 0$^{\circ}$ is face-on), and radial locations of various zones of the disks and the grain radius, $a$, are fit
to the data as described in the text.
\\
Notes: (a) Koester et al. (2005), (b)  Wolff et al. (2002),  (c) Koester \& Wilken (2006)
\begin{figure}
\plotone{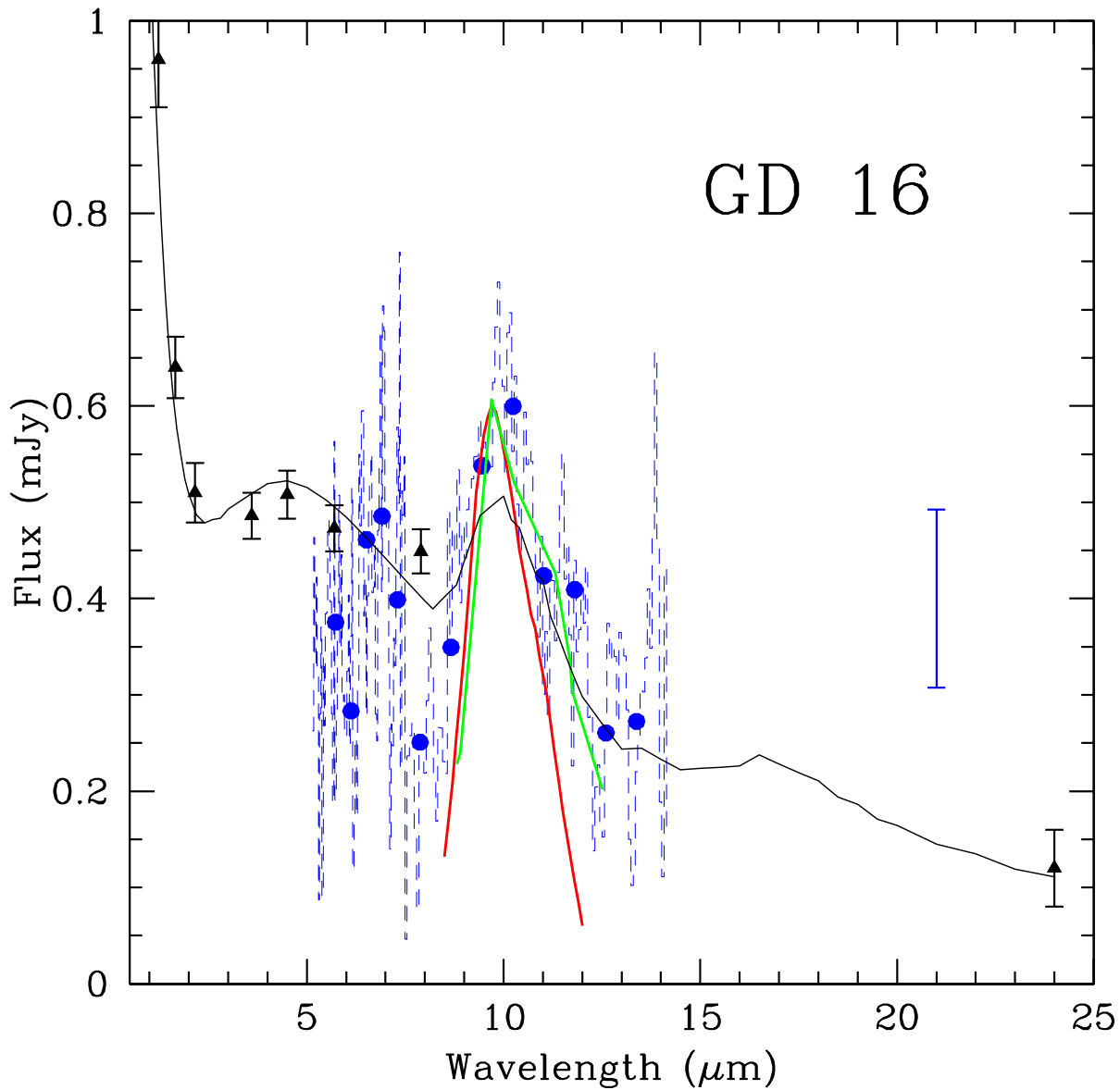}
\caption{Data and model for GD 16.  The unsmoothed data are shown as the dashed blue line.  Smoothing is achieved by  averaging the flux in 13 adjacent pixels and each point is shown as a filled blue circle. For clarity of presentation of the IRS data, we show only one offset error bar which is derived from the standard deviation  of the smoothed data in the 6-8 ${\mu}$m region.   Broad-band IRAC and MIPS fluxes
are shown as black triangles with 1${\sigma}$ error bars (see Jura et al. 2007a, Farihi et al. 2008a,b); the $J$, $H$ and $K$-band data are from Farihi et al. (2008b).  The red line is the interstellar silicate feature of Kemper et al. (2004) inverted from absorption to emission and normalized to have a maximum equal to the highest smoothed data point.  The green line is the profile of the silicate emission in BD +20 307 (Song et al. 2005) and similarly normalized.  The black curve represents the flat disk model described in the text.}
\end{figure}
\begin{figure}
\plotone{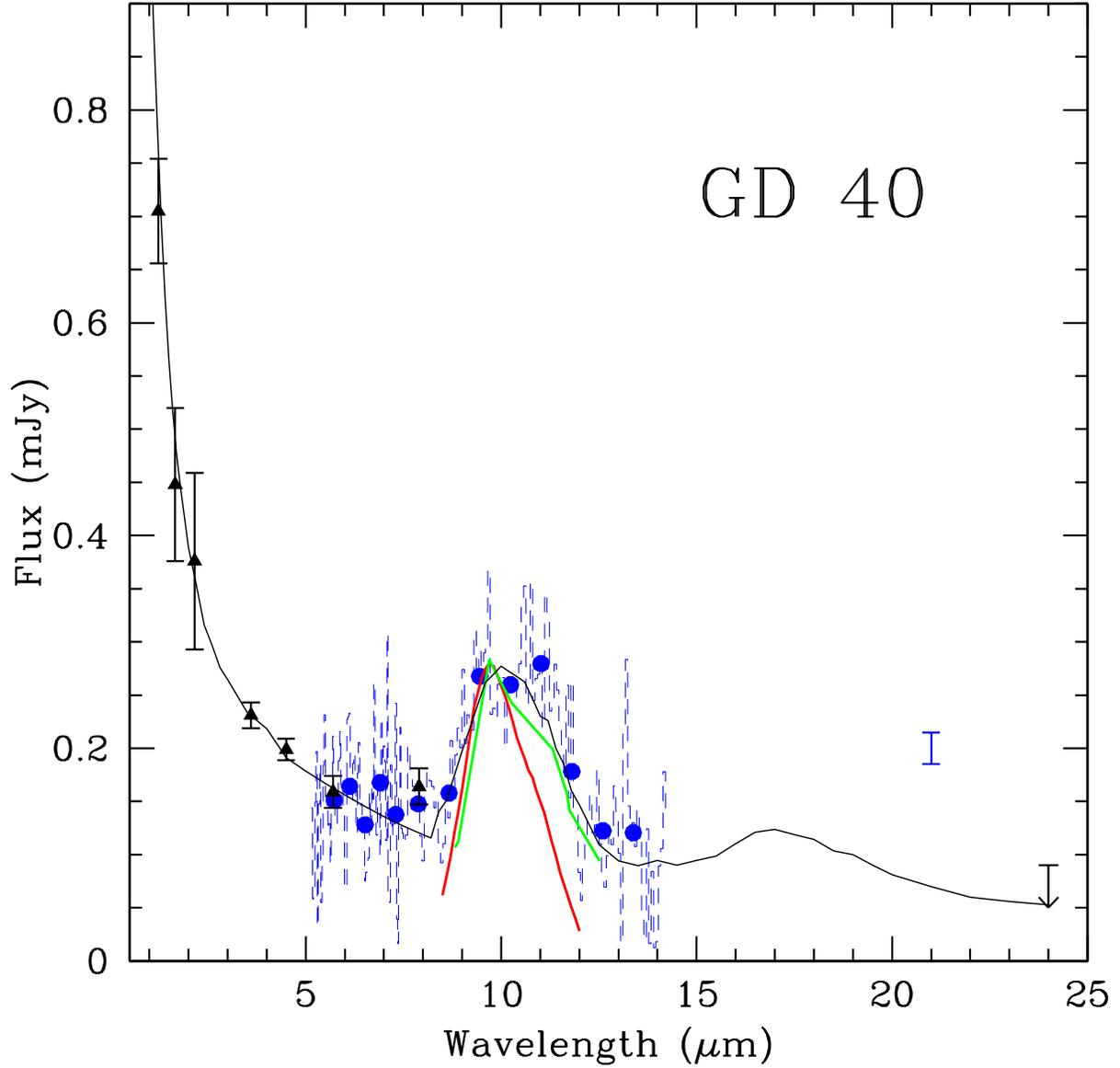}
\caption{Same as Figure 1 except for GD 40.  The $J$, $H$ and $K_{s}$-band data are from 2MASS;  the upper limit at 24 ${\mu}$m is from MIPS photometry.}
\end{figure}
\begin{figure}
\plotone{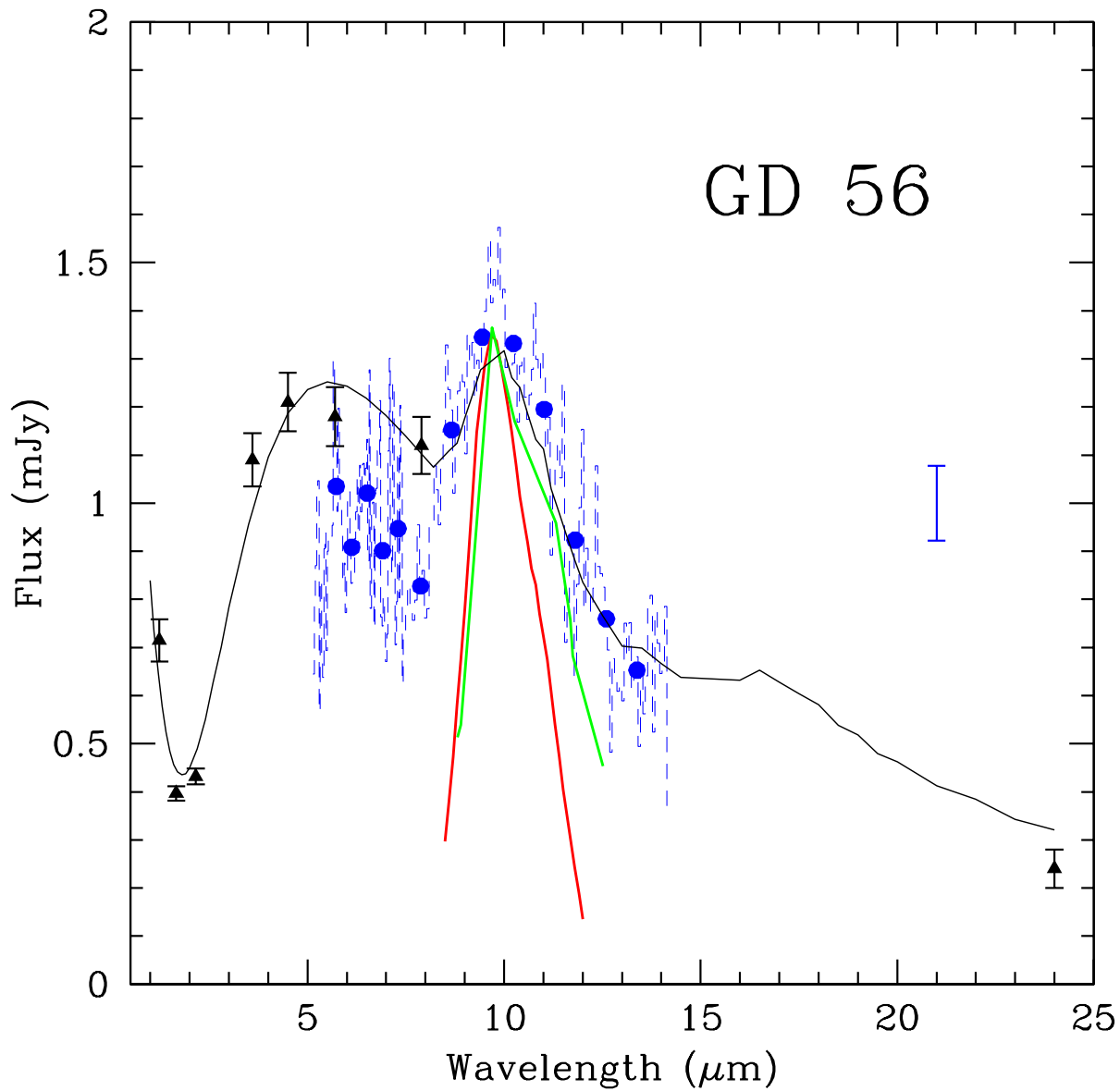}
\caption{Same as Figure 1 except for GD 56, however the $J$, $H$ and $K$-band data are from Kilic et al. (2006).  Also, we assume a warped disk model as described in ${\S}$3.    The inner radius of the warp is at 35 $R_{*}$ and a cosine of the warp angle, as defined in the text,  of 0.075 is assumed.}
\end{figure}
\begin{figure}
\plotone{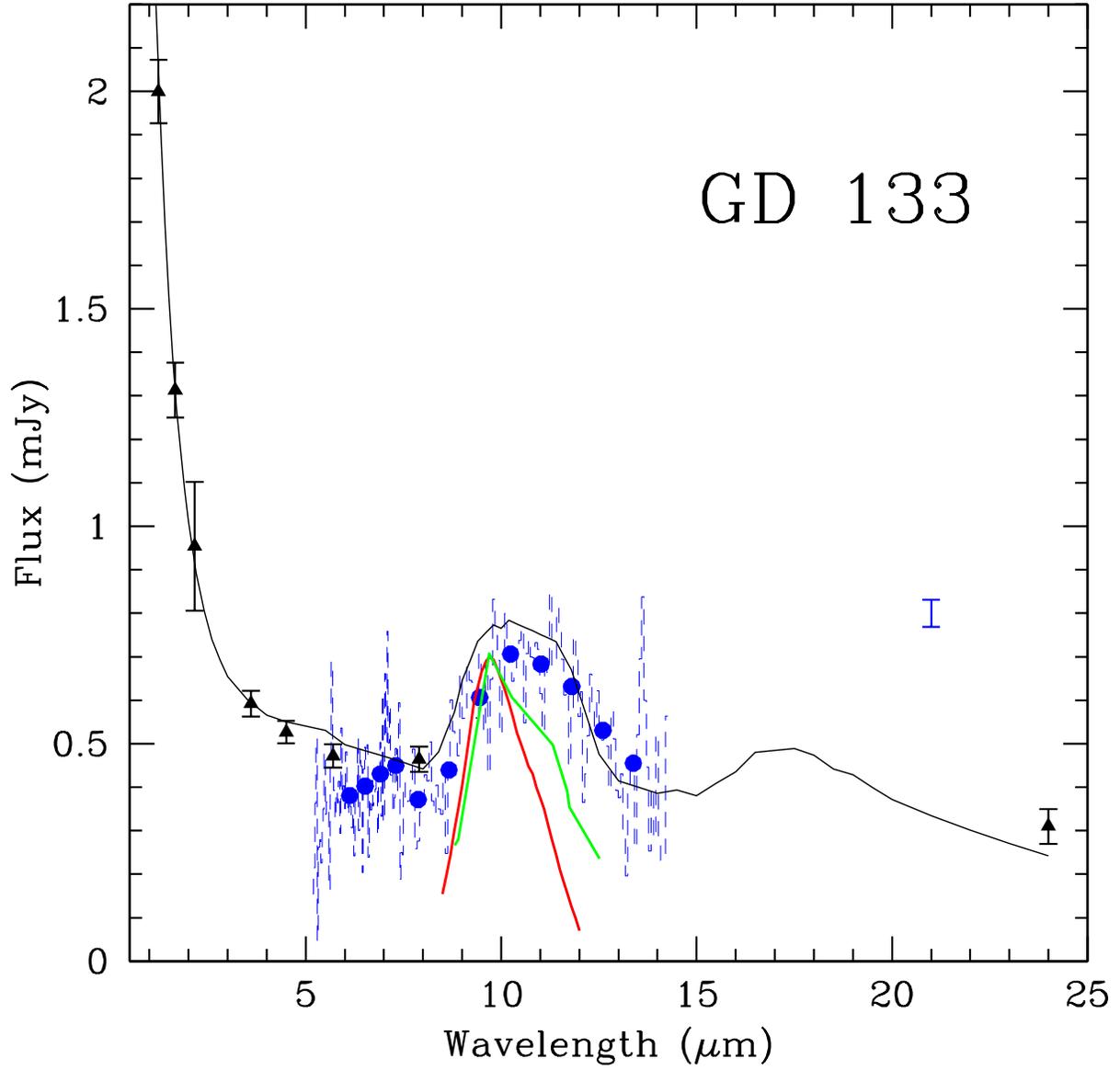}
\caption{Same as Figure 1 except for GD 133 and the $J$, $H$ and $K_{s}$-band data are from 2MASS.}
\end{figure}
\begin{figure}
\plotone{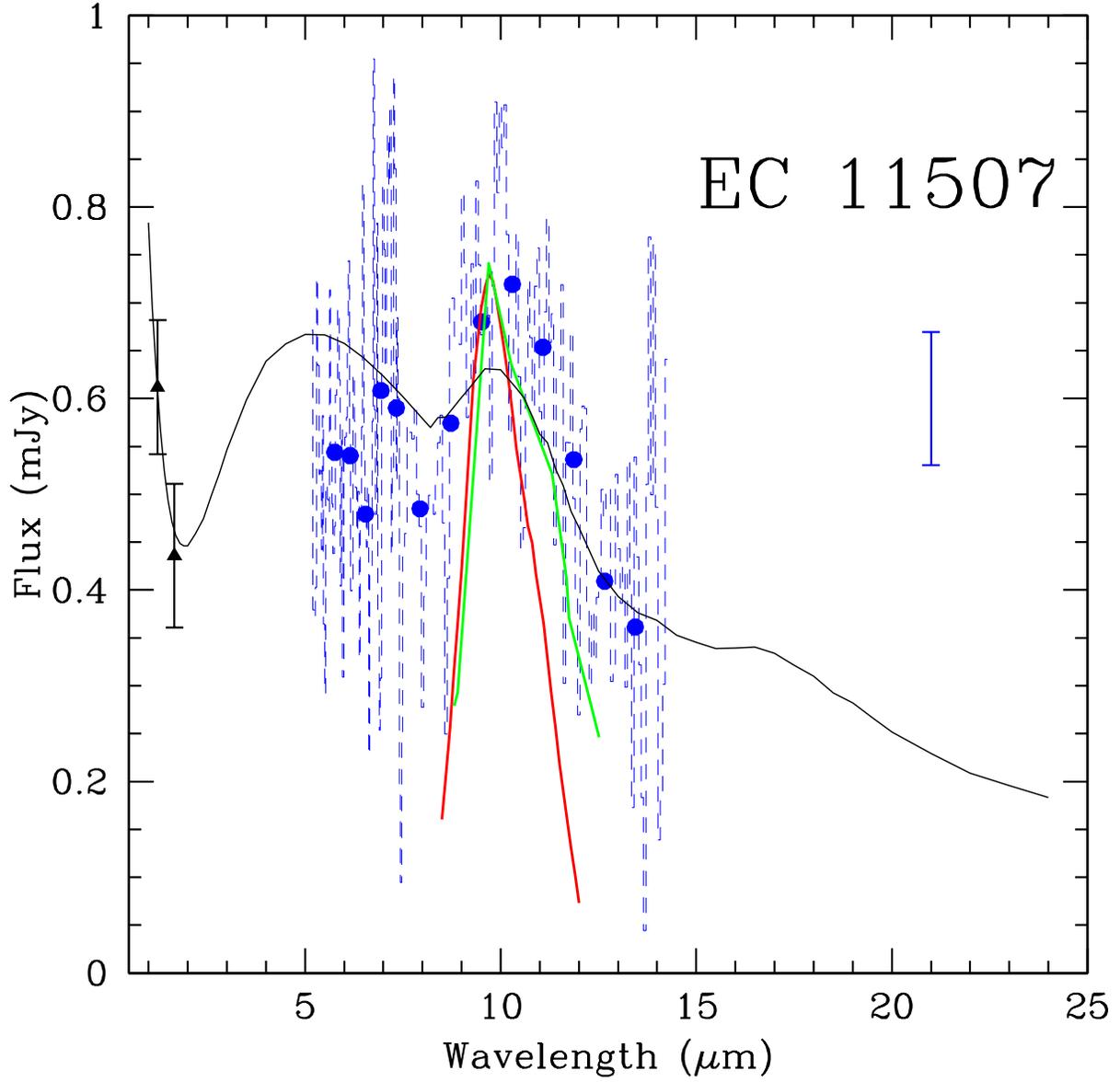}
\caption{Same as Figure 1 except for EC 11507$-$1519 and the $J$, $H$-band data are from 2MASS.}
\end{figure}
\begin{figure}
\plotone{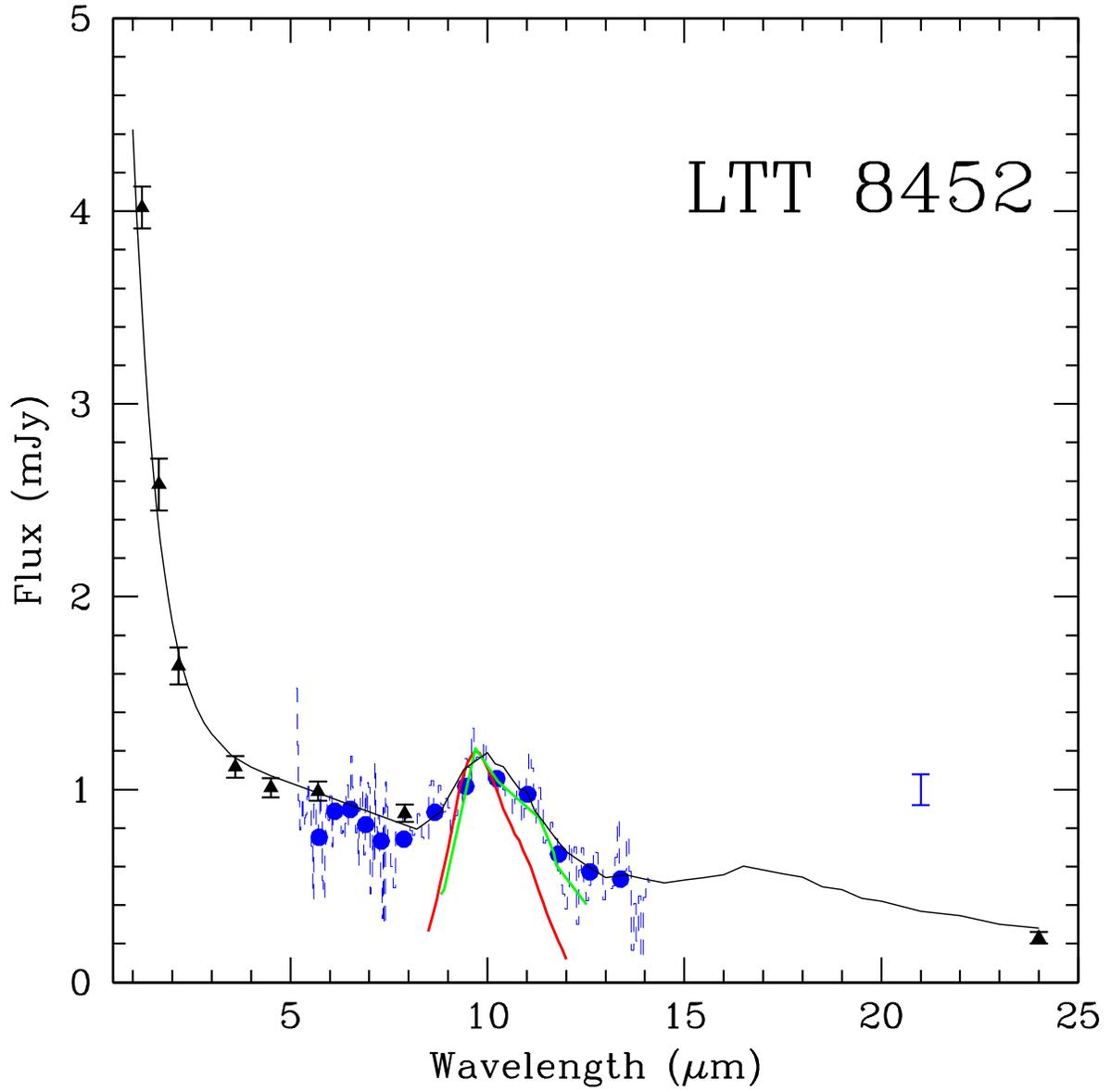}
\caption{Same as Figure 1 except for LTT 8452 and the $J$, $H$, and $K_{s}$-band data are from 2MASS. As discussed ${\S}$2, the IRS spectra have been
scaled up by a factor of 2.5 to match the fluxes derived from IRAC.}
\end{figure}
\begin{figure}
\plotone{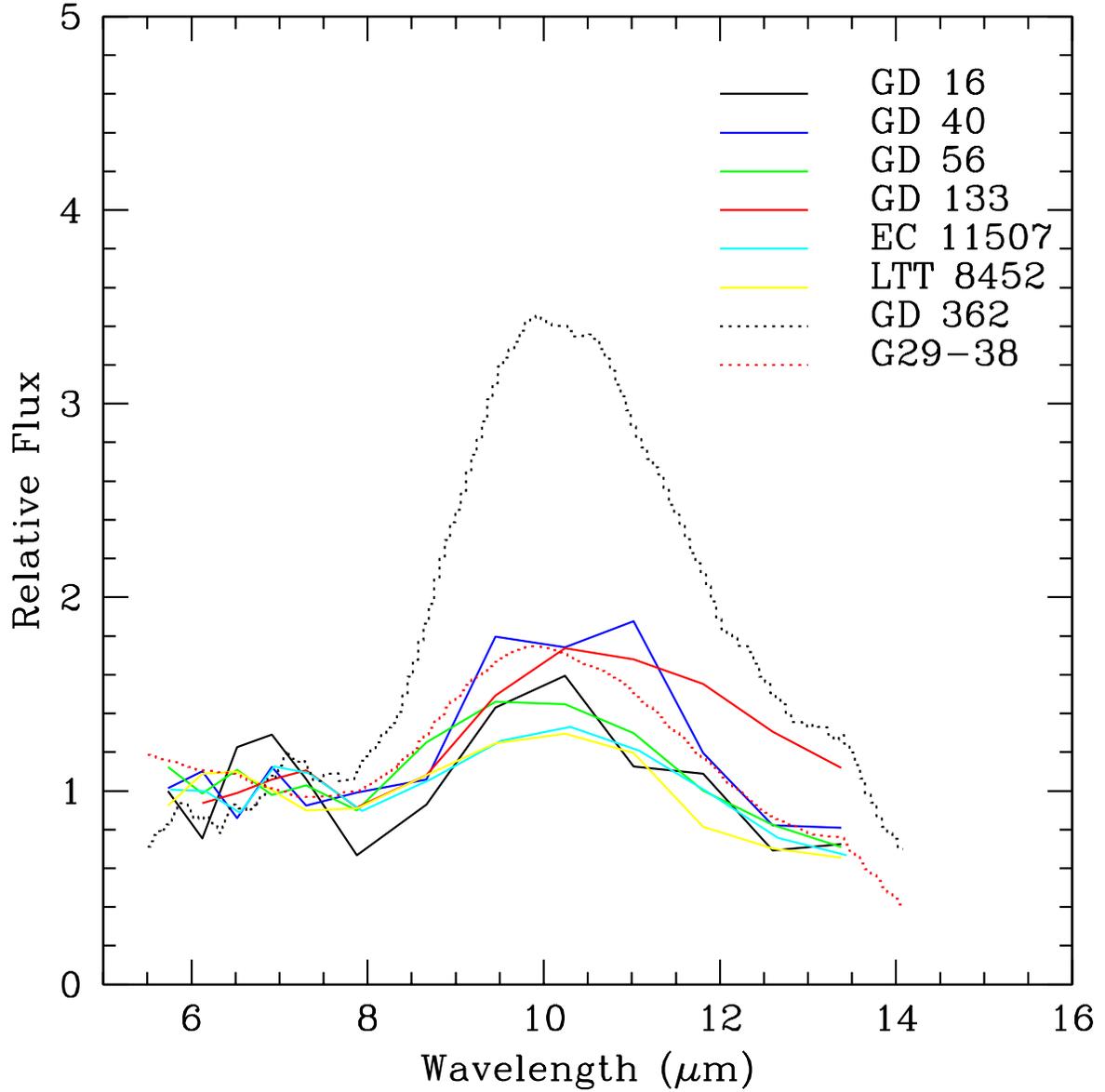}
\caption{Relative smoothed spectra of white dwarfs with infrared excess emission  normalized
to the average flux between 6 ${\mu}$m and 8 ${\mu}$m.  The solid curves represent data  from this paper; the dotted curves represent archived data for  G29-38 (Reach et al. 2005) and GD 362 (Jura et al. 2007b).}
\end{figure}
\end{document}